\documentclass[preprint,aps,12pt,preprintnumbers,eqsecnum,nofootinbib]{revtex4}
\usepackage{graphicx}
\usepackage{subfigure}

\usepackage{color}
\usepackage{amssymb,amsmath}

\newcommand{\beq}{\begin{equation}}
\newcommand{\enq}{\end{equation}}

\begin{document}
% 
%
% Title of paper
\title{\vspace*{0.5in} 
LHC Constraints on the Lee-Wick Higgs Sector
\vskip 0.1in}
\author{Christopher D. Carone}\email[]{cdcaro@wm.edu}
\author{Raymundo Ramos}\email[]{raramos@email.wm.edu}
\author{Marc Sher}\email[]{mtsher@wm.edu}

\affiliation{High Energy Theory Group, Department of Physics,
College of William and Mary, Williamsburg, VA 23187-8795}
\date{February 2014}

\begin{abstract}
We determine constraints on the Lee-Wick Higgs sector obtained from the full LHC Higgs boson data set.  We 
determine the current lower bound on the heavy neutral Lee-Wick scalar, as well as projected bounds at a 
14 TeV LHC with 300 and 3000 inverse femtobarns of integrated luminosity.  We point out that the first sign
of new physics in this model may be the observation of a deviation from standard model expectations
of the lighter neutral Higgs signal strengths corresponding to production via gluon-gluon fusion and 
decay to either tau or $Z$ pairs.   The signal strength of the latter is greater than the standard model expectation, unlike most
extensions of the standard model.
\end{abstract}
\pacs{}
\maketitle

\section{Introduction}\label{sec:intro}
Over the past three decades, the most popular approach to addressing the hierarchy problem of the standard model
has been to introduce additional particles whose virtual effects lead to a cancellation of quadratic divergences.  Supersymmetry 
has been the most studied scenario of this type; only a few years ago, there was much anticipation that colored superparticles 
would be revealed early in the first run of the Large Hadron Collider (LHC).  Unfortunately, this expectation has not been realized.
Since theories with partner particles have a decoupling limit, it is possible that the colored partners, which the LHC is most capable of 
detecting, may lie just beyond the reach of the initial $\sim$8 TeV run.  It also follows that alternatives to supersymmetry, with their 
own distinct set of partner particles, remain in play as possible solutions to the hierarchy problem.  Here we determine how effectively current 
LHC data on the Higgs boson can constrain one such possibility, and explore the reach attainable in the future.

We assume the framework of the Lee-Wick Standard Model (LWSM)~\cite{lwsm}.  In the LWSM, a higher-derivative term 
quadratic in the fields is introduced for each standard model particle.  An additional pole in each propagator corresponds to 
a new physical state, the Lee-Wick partner.  Quadratic divergences in the theory are eliminated due to the faster fall-off of the 
momentum-space propagators in the higher-derivative formulation of the theory.  The presence of twice as many time
derivatives in the theory implies that twice as much initial-value data is needed to specify solutions to the classical equations of motion.  
Hence, one anticipates that the theory can be reformulated in terms of an equivalent one with twice as many fields, but kinetic 
terms with only two derivatives. This is precisely what happens in the auxiliary-field formulation of the LWSM~\cite{lwsm}, as we will
illustrate in the next section.  The additional field corresponds to the Lee-Wick partner particle, and the elimination of quadratic
divergences emerges via cancellations between diagrams involving ordinary and Lee-Wick particles, respectively~\cite{lwsm}.

The LWSM is unusual in that the Lee-Wick partner fields have wrong-sign quadratic terms; this implies that the Lee-Wick states
have negative norm.  In the original papers of Lee and Wick~\cite{lw}, as well as Cutkosky {\em et al.}~\cite{clop}, it was argued that the 
unitarity of such a theory could be maintained provided that the Lee-Wick partners are unstable ({\em i.e.,} are excluded from the set 
of possible asymptotic scattering states) and that a specific pole prescription is used in evaluating loop diagrams.   This approach
has proven effective at the level it has been checked (one loop) and it is generally taken as a working assumption that some viable 
prescription exists at higher order.   While Lee-Wick theories violate causality at a microscopic level, it has been argued that this may not lead to 
logical paradoxes~\cite{coleman}.  In the context of scattering experiments, this has been supported by a study of the large-$N$ limit of the 
Lee-Wick O($N$) model, where the unitarity and Lorentz-invariance of the S-matrix could be explicitly confirmed~\cite{emergent}.  While the 
phenomenological implications of microscopic acausality are of substantial interest~\cite{apheno}, they will not be the subject of this paper.
Other phenomenological studies of Lee-Wick theories can be found in Ref.~\cite{otherlw}. 

We focus instead on how the most current LHC data constrains the possibility of Lee-Wick partners.  Specifically, we focus on a 
Lee-Wick extension of the Higgs sector, an effective theory in which the Lee-Wick partner to the Higgs doublet is retained, while all 
the other Lee-Wick partners are assumed to be heavy and decoupled~\cite{carprim,zurita}.  This approximation is justified for the following  
reason:  the Lee-Wick partners to the Higgs field, the electroweak gauge bosons and the top quark are the most important in the cancellation of 
quadratic divergences; these would be expected to be the lightest to minimize fine tuning.  Of this set, however, all but the partner to the 
Higgs doublet are forced up to multi-TeV energy scales by existing electroweak constraints~\cite{lwew}.  As we will show in the next section, the 
Lee-Wick Higgs sector presents itself as an unusual, constrained two-Higgs doublet model, one that is specified by a single free 
parameter once the lightest scalar mass eigenvalue is fixed.   Current data on the 125 GeV Higgs boson at the LHC can then be 
used to determine bounds on the masses of the other neutral and charged scalar mass eigenstates in the theory.  We note that
past studies of the Lee-Wick Higgs sector~\cite{carprim,zurita,figy} were undertaken before LHC Higgs boson data was available; in this
letter we take into account all such data available to date and determine projected bounds based on current assessments
of the integrated luminosities that may be realistically obtained.

Our letter is organized as follows: In Section~\ref{sec:lwhs}, we define our effective theory.  In Section~\ref{sec:bounds}, we determine 
bounds on the heavier neutral scalar by fitting the model's predictions for the $125$~GeV mass eigenstate, using the full data set 
currently available from the LHC.  In the second part of this section, we determine projected  bounds based on the assumption 
of $300$ to $3000$ fb$^{-1}$ of integrated luminosity at a 14~TeV LHC.  In Section~\ref{sec:concl}, we summarize our results
and compare them to other existing bounds on the model.

\section{The Lee-Wick Higgs Sector} \label{sec:lwhs}
In the Higgs sector of our model, a higher-derivative kinetic term is included in the Higgs field Lagrangian
\begin{equation}
{\cal L} = (D_\mu \hat{H})^\dagger (D^\mu \hat{H}) 
- \frac{1}{m_{\tilde{h}}^2} (D_\mu D^\mu \hat{H})^\dagger (D_\nu D^\nu \hat{H}) - V(\hat{H}) \,\, .
\label{eq:lwhhd}
\end{equation}
Here $D_\mu=\partial_\mu - i g W^a_\mu T^a - i g' B_\mu Y$ is the usual covariant derivative for the standard model 
gauge group and a hat denotes a field in the higher-derivative formulation of the theory.   The Higgs potential is given by
\begin{equation}
V(\hat{H}) = \frac{\lambda}{4} \left( \hat{H}^\dagger \hat{H} - \frac{v^2}{2} \right)^2  \,\, .
\end{equation}
Eq.~(\ref{eq:lwhhd}) is reproduced from the following Lagrangian, 
\begin{equation}
{\cal L}= (D_\mu \hat{H})^\dagger (D^\mu \hat{H}) + [(D_\mu \hat{H})^\dagger (D^\mu \tilde{H}) + h.c.]
+ m_{\tilde{h}}^2 \tilde{H}^\dagger \tilde{H} - V(\hat{H}),
\label{eq:auxlag}
\end{equation}
if one eliminates the auxiliary field $\tilde{H}$ using its equation of motion.  If instead, one uses the field redefinition
$\hat{H}=H-\tilde{H}$, Eq.~(\ref{eq:auxlag}) takes the standard Lee-Wick form
\begin{equation}
{\cal L}_{LW}=(D_\mu H)^\dagger (D^\mu H) - (D_\mu \tilde{H})^\dagger (D^\mu \tilde{H}) + m_{\tilde{h}}^2 \tilde{H}^\dagger \tilde{H}
-V(H-\tilde{H}) \, .
\label{eq:lwform}
\end{equation}
In unitary gauge, the Higgs doublet can be decomposed
\begin{equation}
H = \left(\begin{array}{c} 0 \\ \frac{v+h}{\sqrt{2}} \end{array}\right) \, \,\,\,\,\,\,\,\,\,\, 
\tilde{H} = \left(\begin{array}{c} \tilde{h}^+ \\ \frac{\tilde{h} + i \tilde{P}}{\sqrt{2}} \end{array}\right) \,,
\label{eq:ug}
\end{equation}
where $v\approx 246$~GeV is the electroweak scale.  Expanding the potential in terms of its quadratic, cubic and
quartic parts, we find:
\begin{equation}
V^{(2)} = \frac{\lambda \, v^2}{4} (h-\tilde{h})^2 - \frac{m_{\tilde{h}}^2}{2} (\tilde{h}^2 + \tilde{P}^2 + 2 \tilde{h}^+ \tilde{h}^-) \, ,
\label{eq:v2}
\end{equation}
\begin{equation}
V^{(3)} = \frac{\lambda \, v}{4}  (h-\tilde{h}) \left[ (h-\tilde{h})^2 + \tilde{P}^2 + 2 \tilde{h}^- \tilde{h}^+\right] \, ,
\label{eq:v3}
\end{equation}
\begin{equation}
V^{(4)} = \frac{\lambda}{16} \left[ (h-\tilde{h})^2 + \tilde{P}^2 + 2 \tilde{h}^- \tilde{h}^+\right]^2  \,.
\label{eq:v4}
\end{equation}
Note that the Lee-Wick charged scalar and pseudoscalar Higgs fields have mass $m_{\tilde{h}}$, while there is
mixing between the neutral scalar states $h$ and $\tilde{h}$.  Indicating the neutral mass eigenstates with the subscript $0$,
we define the mixing angle
\begin{equation}
\left(\begin{array}{c} h \\ \tilde{h} \end{array} \right) = \left(\begin{array}{cc} \cosh\alpha & \sinh\alpha \\ \sinh\alpha & \cosh\alpha \end{array} \right) \left(\begin{array}{c} h_0 \\ \tilde{h}_0 \end{array} \right) \, .
\label{eq:rot}
\end{equation}
The symplectic rotation is necessary to preserve the relative sign between the ordinary and Lee-Wick kinetic terms. It follows from 
Eq.~(\ref{eq:v2}) that
\begin{equation}
\tanh 2 \alpha = -\frac{2 m_h^2/m_{\tilde{h}}^2}{1-2 m_h^2/m_{\tilde{h}}^2} \,\,\,\,\, \mbox{ or } \,\,\,\,\,  \tanh \alpha = - m_{h_0}^2 / m_{\tilde{h}_0}^2  \,\, ,
\label{eq:tanha}
\end{equation}
where $m_h^2 \equiv \lambda \, v^2 /2$ is the mass of the lighter Higgs scalar in the absence of mixing.  The mass 
squared eigenvalues are defined by $m_{h_0}^2$ and $-m_{\tilde{h}_0}^2$, so that the squared mass parameters 
appearing in Eq.~(\ref{eq:tanha}) are all positive.   Note that $\alpha$ is always negative.

The same steps that led to Eq.~(\ref{eq:lwform}) determine the form of the Yukawa couplings
\begin{eqnarray}
{\cal L}&=&\frac{\sqrt{2}}{v}\,  \overline{u}_R \, m_u^{diag} (H-\tilde{H}) i \sigma^2 Q_L
-\frac{\sqrt{2}}{v} \, \overline{d}_R \, m_d^{diag} (H-\tilde{H})^\dagger V^\dagger_{\rm CKM} Q_L \nonumber \\
&-& \frac{\sqrt{2}}{v} \, \overline{e}_R \, m_e^{diag} (H-\tilde{H})^\dagger \ell_L + \mbox{ h.c.},
\label{eq:yuk}
\end{eqnarray}
where we have suppressed generation indices.  Here  $Q_L \equiv ( u_L \, , \,V_{\rm CKM} d_L)$, $\ell_L \equiv (\nu_L \, , \, e_L)$, 
and all the fermion fields shown are in the mass eigenstate basis.  The couplings of the neutral scalar mass eigenstates 
to fermions can now easily be extracted using Eqs.~(\ref{eq:ug}) and (\ref{eq:rot}).

We define the quantity $g_{XY}$ to be the ratio of a neutral scalar coupling in the Lee-Wick theory that we have defined to the 
same coupling of the Higgs boson in the standard model. Here  $X$ designates the scalar state (either $h_0$ or $\tilde{h}_0$) 
and $Y$ specifies the coupling of interest (for example, $t\overline{t}$, $b\overline{b}$, $\tau^+\tau^-$, $W^+W^-$ or $ZZ$).  The 
neutral Higgs couplings to gauge boson pairs can be extracted from Eq.~(\ref{eq:lwform}) and the couplings to fermions 
from Eq.~(\ref{eq:yuk}).   For example, we find
\begin{equation}
g_{h_0 t \overline{t}} = g_{h_0 b \overline{b}}=g_{h_0 \tau \tau} = e^{-\alpha} \,\,\, ,
\end{equation}
\begin{equation}
g_{h_0 WW} = g_{h_0 ZZ} = \cosh\alpha \,\,\,\, ,
\end{equation}
\begin{equation}
g_{\tilde{h}_0 t \overline{t}} = g_{\tilde{h}_0 b \overline{b}}=g_{\tilde{h}_0 \tau \tau} = - e^{-\alpha} \,\,\, ,
\end{equation}
\begin{equation}
g_{\tilde{h}_0 WW} = g_{\tilde{h}_0 ZZ} = \sinh\alpha \,\,\,\, .
\end{equation}
Note that the couplings $g_{h_0 WW}$ and $g_{h_0 ZZ}$ are bigger than one, unlike most extensions of the standard model.  
These results provide most of what we need to modify known theoretical results for Higgs boson properties in
the standard model to obtain those appropriate to the scalar states in the present theory.   The one coupling that is more 
complicated to modify is the effective Higgs coupling to two photons; the relevant one-loop amplitude depends on 
a sum of terms that are modified by different $\alpha$-dependent factors.   To proceed, we write
the relevant Lee-Wick Lagrangian terms as
\begin{eqnarray}
{\cal L} &=& -\frac{g \, m_f}{2 m_W} e^{-\alpha} (h_0 - \tilde{h}_0) \overline{f} f
+  (\cosh\alpha \, h_0 + \sinh\alpha \, \tilde{h}_0) \, g  \, m_W W^+ W^- \nonumber \\  
&& - \left(\frac{1}{2} \frac{m_h^2}{m_{\tilde{h}}^2} e^{-\alpha} \right)  \frac{g \, m_{\tilde{h}}^2}{m_W} (h_0 - \tilde{h}_0) \,  
\tilde{h}^- \tilde{h}^+  \,\,\, .
\end{eqnarray}
Presented in this form, coefficients can be easily matched to those of the effective Lagrangian assumed in Ref.~\cite{hhg} to 
compute contributions to $h_0 \rightarrow \gamma \gamma$ from intermediate loop particles of various spins.
After identifying the appropriate coupling factors, the only other modification that needs to be made to these
generic formulae is that an additional minus sign must be included in the amplitude term corresponding
to the charged Higgs loop;  this takes into account the overall sign difference between ordinary and Lee-Wick 
propagators.

\begin{table}[tp]
\caption{Measured Higgs Signal Strengths}
\centering
\begin{tabular}{cccc}
\hline\hline
 Decay            & Production & Measured Signal Strength $R^{meas}$ \\ 
\hline
{$\gamma \gamma$} & ggF+tth    & $1.6^{+ 0.3 +0.3}_{-0.3-0.2}$, [ATLAS] \cite{atlas-13012}\\ 
                  & VBF        & $1.9^{+ 0.8}_{-0.6}$  [ATLAS]\cite{Aad:2013wqa}\\ 
                  & Vh         & $1.3^{+ 1.2}_{-1.1}$  [ATLAS]\cite{Aad:2013wqa}\\ 
                  & inclusive  & $1.55^{+ 0.33}_{-0.28}$   [ATLAS]\cite{Aad:2013wqa}\\ 
                  & ggF+tth    & $0.52 \pm 0.5$   [CMS]\cite{cms-13001}\\ 
                  & VBF+Vh     & $1.48^{+1.24}_{-1.07}$  [CMS]\cite{cms-13001}\\ 
                  & Inclusive  & $0.78^{+0.28}_{-0.26}$      [CMS]\cite{cms-13001}\\
                  & ggF        & $6.1^{+3.3}_{-3.2}$  [Tevatron]\cite{hcp:Enari}\\
                  \hline
{$W W$}       & ggF        & $0.82^{+0.33}_{-0.32}$          [ATLAS] \cite{Aad:2013wqa}\\ 
              & VBF        & $1.4^{+0.7}_{-0.6}$    [ATLAS]\cite{Aad:2013wqa}\\
              & VBF+Vh     & $1.66 \pm 0.79$    [ATLAS] \cite{atlas-13030}\\ 
              & Inclusive  & $0.99^{+0.31}_{-0.28}$   [ATLAS]\cite{Aad:2013wqa}\\ 
              & ggF        & $0.76 \pm 0.21$        [CMS]\cite{cms-13003}\\ 
              & ggF+VBF+Vh & $0.72^{+0.20}_{-0.18}$        [CMS]\cite{Chatrchyan:2013iaa}\\ 
              & ggF        & $0.8^{+0.9}_{-0.8}$    [Tevatron]\cite{hcp:Enari}\\
              \hline
{$ZZ$}      & ggF+tth    & $1.45^{+0.43}_{-0.36}$   [ATLAS] \cite{Aad:2013wqa}\\ 
            & VBF+Vh     & $1.2^{+1.6}_{-0.9}$   [ATLAS]\cite{Aad:2013wqa}\\ 
            & Inclusive  & $1.43^{+0.40}_{-0.35}$         [ATLAS]\cite{Aad:2013wqa}\\ 
            & ggF        & $0.9^{+0.5}_{-0.4}$   [CMS] \cite{cms-13002}\\ 
            & VBF+Vh     & $1.0^{+2.4}_{-2.3}$   [CMS]\cite{cms-13002}\\ 
            & inclusive  & $0.93^{+0.26+0.13}_{-0.23-0.09}$ [CMS]\cite{Chatrchyan:2013mxa}\\
\hline\hline
\end{tabular}
\label{tboson}
\end{table}

\section{Bounds}~\label{sec:bounds}
The quantities that we compute for purpose of comparison to the experimental data are the signal strengths
$R^{\rm LW}_i$, each a specified Higgs boson production cross section times branching fraction normalized to the standard
model expectation for the same quantity.  We consider production via gluon-gluon fusion (ggF), vector-boson fusion (VBF),
associated production with a W or Z boson (Vh) and production via the top quark coupling (tth), as well as combinations of 
these possibilities.  In most cases, the ratio of Lee-Wick to standard model Higgs production cross sections reduces to a simple 
factor (for example, $e^{-2\alpha}$ for ggF).  In the case of inclusive production at the LHC, we find that the ratio is well approximated
by
\begin{equation}
\frac{\sigma^{\rm LW}}{\sigma^{\rm SM} }= 0.88 \, e^{-2 \alpha} + 0.12 \, \cosh^2\alpha \, ,
\end{equation}
for a center-of-mass energy of either 8 or 14~TeV.  The coefficients in this expression were determined using
numerical predictions for the different contributions to the standard model Higgs production cross section, given in Ref.~\cite{cernXs}.

A total of 33 signal strengths measured at ATLAS, CMS and the Tevatron were collected for analysis; they correspond to different 
channels of Higgs production and decay, and include the final states $\gamma\gamma$, $ZZ$, $WW$, $bb$ and $\tau\tau$ 
(Tables~\ref{tboson} and \ref{tfermion}). The analysis performed here is analogous to others found in  the 
literature~\cite{Chen:2013kt,sherdaw,haothers}.  These references considered
conventional two-Higgs doublet models, with results plotted as a function of $\alpha$ and $\tan\beta$.  We have seen,
however, that the Lee-Wick Higgs sector is determined by a single parameter $\alpha$;  as indicated by Eq.~(\ref{eq:tanha}),
this mixing angle is in one-to-one correspondence with the value of the heavy scalar mass $m_{\tilde{h}_0}$ after one fixes 
$m_{h_0}$ at its experimental value.  Hence, we will present our results as 95\% C.L. lower bounds on 
the heavy Lee-Wick scalar mass.

This analysis is presented in two parts: We first determine bounds using the most recent data for the Higgs boson signal strengths shown
in Tables~\ref{tboson} and \ref{tfermion}.  We then determine projected bounds at a 14~TeV LHC by assuming that the experimental data will 
converge on standard model central values and that the errors will scale in a simple way with the integrated luminosity.

\begin{table}[tp]
\caption{Measured Higgs Signal Strengths}
\centering
\begin{tabular}{cccc}
\hline\hline
 Decay          & Production & Measured Signal Strength $R^{meas}$ \\
\hline
{$b\bar{b}$}    & Vh         & $0.2 \pm0.5 \pm0.4$     [ATLAS] \cite{TheATLAScollaboration:2013lia}\\ 
                & Vh         & $1.0 \pm0.5$            [CMS]\cite{Chatrchyan:2013zna}\\ 
                & Vh         & $1.56^{+0.72}_{-0.73}$  [Tevatron]\cite{hcp:Enari}\\
\hline
{$\tau^+ \tau^-$} & ggF        & $1.1^{+1.3}_{-1.0}$    [ATLAS]\cite{atlas-108}\\ 
                  & VBF        & $-0.4 \pm1.5$          [ATLAS]\cite{atlas-160}\\ 
                  & VBF+Vh     & $1.6^{+0.8}_{-0.7}$    [ATLAS]\cite{atlas-108}\\ 
                  & ggF+VBF+Vh & $1.4^{+0.5}_{-0.4}$    [ATLAS]\cite{atlas-108}\\ 
                  & ggF        & $0.73 \pm 0.50$        [CMS]\cite{cms-13004}\\ 
                  & VBF        & $1.37^{+0.56}_{-0.58}$ [CMS]\cite{cms-13004}\\ 
                  & Vh         & $0.75^{+1.44}_{-1.40}$ [CMS]\cite{cms-13004}\\ 
                  & Inclusive  & $0.78 \pm 0.27$        [CMS]\cite{Chatrchyan:2014nva}\\ 
                  & ggF        & $2.1^{+2.2}_{-1.9}$    [Tevatron]\cite{hcp:Enari}\\
\hline\hline
\end{tabular}
\label{tfermion}
\end{table}

To find a lower bound on $m_{\tilde{h}_0}$ from the current signal strengths, we construct the $\chi^2$ function
\begin{equation}
\chi^2=\sum_{i=1}^{33}\left(\frac{R^{\text{LW}}_i-R_i^{\text{meas}}}{\sigma^{\text{meas}}_i}\right)^2,
\label{eq:chisq}
\end{equation}
where $i$ runs over the 33 channels in Tables~\ref{tboson} and \ref{tfermion}. $R^{\text{LW}}_i$ stands for the predicted strength in the model 
presented here, $R^{\text{meas}}_i$ is the measured strength and $\sigma^{\text{meas}}_i$ is the corresponding error. Asymmetric errors were 
averaged in quadrature,  $\sigma=\sqrt{(\sigma_+^2+\sigma_-^2)/2}$.  Note that only experimental errors were taken into account; in most cases,
theoretical errors cancel in the ratio of a given observable with its standard model expectation.  In the cases where the cancellation is not exact, the
theoretical uncertainty remains small.  For example, an ${\cal O}(10\%)$ theoretical uncertainty in the ggF production cross section does not entirely 
scale out the ratio of inclusive production cross sections;  however, this translates into an ${\cal O}(1\%)$ theoretical uncertainty in the ratio, which is
much smaller than the current experimental error bars.

We determine the 95\% C.L. lower bound on the heavy neutral Lee-Wick scalar mass using Eq.~(\ref{eq:chisq}) and the $\chi^2$ probability distribution corresponding to $32$ degrees of freedom.   For the data in Tables~\ref{tboson} and \ref{tfermion}, which includes $\sim25$~fb$^{-1}$
of LHC data at $\sim 8$~TeV, we find 
\begin{equation}
m_{\tilde{h}_0}>255\mbox{ GeV}  \,\,\,\,\, \mbox{ 95\% C.L.}
\label{eq:cb}
\end{equation}
This corresponds to a mixing parameter $\alpha \approx -0.25$. 

\begin{figure}[t]
  \begin{center}
    \includegraphics[width=0.5\textwidth]{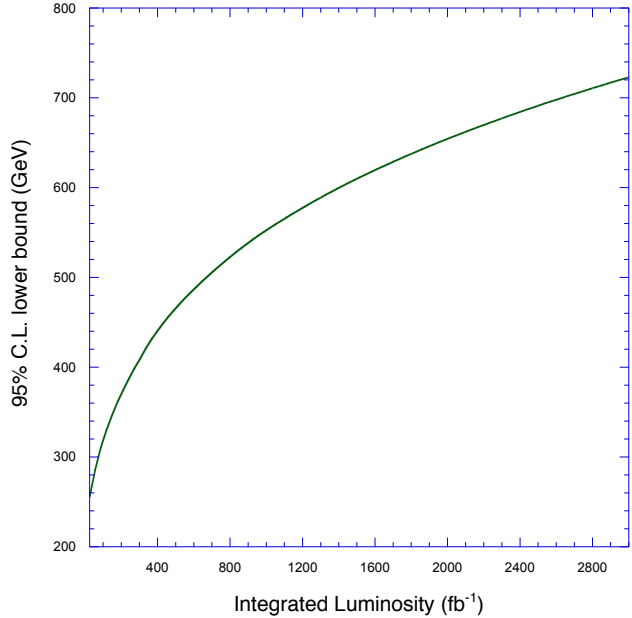}
    \caption{Projected lower bound on the heavy Lee-Wick scalar mass $m_{\tilde{h}_0}$ as a function
    of the LHC integrated luminosity.}
    \label{fig:lowvsILplot}
  \end{center}
\end{figure}

To estimate the future reach of the LHC, we follow the same procedure as in Ref.~\cite{sherdaw}.  We assume that the experimental signal
strengths will converge to their standard model values, namely $R_i^{\rm meas} =1$, and that the experimental errors bars will shrink
relative to their current values by a factor of $1/\sqrt{N}$ where
\begin{equation}
N=\frac{\sigma_{\text{14}}}{\sigma_{\text{8}}}\frac{L_{14}}{L_8}.
\end{equation}
Here, $\sigma_{X}$ is the total Higgs production cross section at center-of-mass energy $X$, and $L_X$ is the corresponding integrated luminosity.
This scaling of errors as the inverse square root of the number of events was also done in Ref. \cite{sherdaw} , and corresponds to ``scheme 2'' of 
the CMS~\cite{cms-12006} high luminosity projections\footnote{The assumption that the uncertainty scales as one over the square root of the 
number of events is true for the statistical error.  Here we assume that a comparable reduction in the systematic errors is possible with increasing 
luminosity.  This assumption may be optimistic, but is the one used by CMS for the European Strategy Report~\cite{cms-12006}.}.

\begin{figure}
  \begin{center}
    \includegraphics[width=0.5\textwidth]{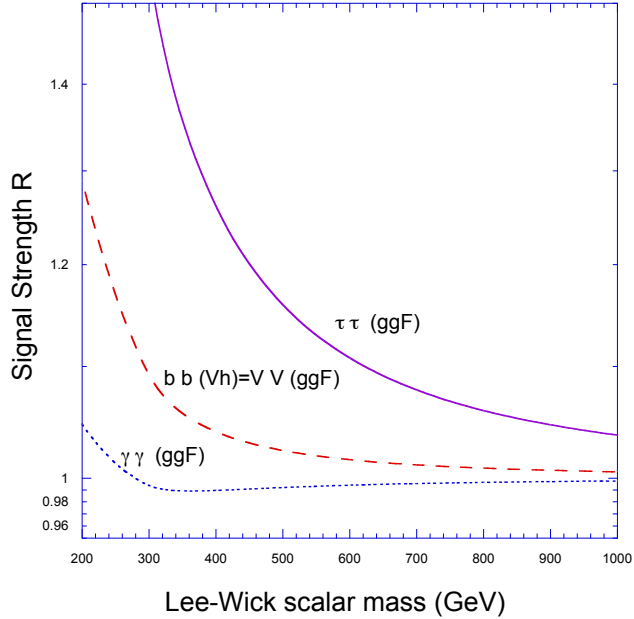}
    \caption{Model predictions for the signal strengths $R_i^{\rm LW}$ as a function of $m_{\tilde{h}_0}$.}
    \label{fig:RLWplot}
  \end{center}
\end{figure}

The results of our projection are shown in Fig.~\ref{fig:lowvsILplot}.  As one might expect, the lower bound on $m_{\tilde{h}_0}$ increases
monotonically with integrated luminosity; the left-most point on the curve corresponds to the current bound in Eq.~(\ref{eq:cb}), while
the rest follow from our procedure for determining projected bounds at a 14~TeV LHC.  For two benchmark points, we find
\begin{align}
m_{\tilde{h}_0}&>420\ \text{GeV} \mbox{ 95\% C.L.}   \,\,\,\,\,   (L_{14} = 300\ \text{fb}^{-1}) \\
m_{\tilde{h}_0}&>720\ \text{GeV} \mbox{ 95\% C.L.}     \,\,\,\,\,  (L_{14} = 3000\ \text{fb}^{-1})
\end{align}
corresponding to the mixing angles $\alpha \approx-0.09$ and $-0.03$, respectively.   We discuss the implications of these bounds in the
final section.

As the experimental uncertainties on the Higgs boson signal strengths become smaller, new physics in this model should become
manifest by an emerging pattern of deviations from the standard model expectations.   To illustrate this, we show in Fig.~\ref{fig:RLWplot}
some of the signal strengths expected in the Lee-Wick theory as a function of the heavy Lee-Wick scalar mass.  The $\tau\tau$ mode
via ggF shows the greatest deviation from the standard model since both the production and decay width are each modified by the 
factor $\exp(-2 \alpha)$ which is larger than one.  The signal strength for $H \to VV$ decays is also enhanced.  Although the deviation is
not as great as the $\tau\tau$ channel shown, there are very few extensions of the standard model that would lead to such an 
enhancement.  Hence, this effect is a distinctive feature of the model that might be identified if the underlying physics is realized in
nature.

\section{Conclusions}~\label{sec:concl}
Now that the LHC has discovered a light, standard model-like Higgs boson and begun a study of its properties, one can examine the current and future constraints 
that can be placed on standard model extensions.   In this letter, we have considered such constraints on a Lee-Wick extension of the Higgs sector.   Although most 
of the partners in the LWSM must be heavy, due to various low-energy constraints, the partners of the Higgs boson need not be.    The resulting effective theory is a 
constrained two-Higgs doublet model, one in which some propagators and
vertices have unusual signs.  In addition, the mixing between the light Higgs and the heavy neutral scalar is described by a symplectic rotation, leading to hyperbolic 
functions of a mixing angle at the vertices. The mixing angle itself is related to the two neutral Higgs masses, and thus the heavy neutral scalar mass can be taken 
as the only free parameter.   The charged and pseudoscalar Higgs masses are degenerate at tree level and are also determined once the heavy scalar mass has 
been specified.

We first considered the bounds from current LHC data, looking at 33 different signals, and found a $95\%$ confidence level lower bound of  
$255$ GeV on the heavy scalar mass.     Extrapolating to the next runs at the LHC (at 14 TeV), we found that the bound will increase to $420$ 
GeV ($720$ GeV) for an integrated luminosity of $300$ ($3000$) inverse femtobarns.    The first signature of a deviation will come from light 
Higgs boson decays to either tau or massive gauge boson pairs.  Unlike most extensions of the standard model, both of these signal strengths are 
greater than in the standard model.

In Ref.~\cite{carprim}, it was shown that flavor constraints on the Lee-Wick charged Higgs provide a lower bound on the heavy neutral scalar mass.  The 95\% C.L. 
bounds on the charged Higgs from $B_d-\bar{B_d}$, $B_s-\bar{B_s}$ mixing and $b\to X_s\gamma$ were found to be $303$ GeV, $354$ GeV and $463$ GeV, 
respectively~\cite{carprim}.   The most stringent of these bounds translates into a lower bound on
the heavy neutral scalar of $445$~GeV.   We thus see that the current bound from $b\to X_s\gamma$ is more stringent than those from 
current Higgs data, and that it will require approximately $400$ femtobarns at a $14$~TeV LHC in order to supersede this bound.

%%%%%%%%%%%%%%%%%%%%%%%%%%%%%%%%%%%%%%%%%%%%%%%%%%%%%%%%%%%
\begin{acknowledgments}  
The work of CC and RR was supported by the NSF under Grant PHY-1068008.  The opinions and conclusions expressed herein are those of the authors, and do not represent the National Science Foundation.
\end{acknowledgments}
%%%%%%%%%%%%%%%%%%%%%%%%%%%%%%%%%%%%%%%%%%%%%%%%%%%%%%%%%%%     

%\appendix
%\section{}

\end{document}